\documentclass[a4paper, amsfonts, amssymb, amsmath]{article}
\usepackage[utf8]{inputenc}
\usepackage{color}
\usepackage[dvipsnames]{xcolor}
\usepackage{xfrac}
\usepackage{cite}

\usepackage{appendix}
\usepackage{soul}
\usepackage{ulem}
\usepackage{graphicx}
\usepackage{soul}
\usepackage{authblk}
\usepackage{amsmath,amssymb,amsfonts,graphicx}
\usepackage[a4paper,top=2cm,bottom=2cm,left=2cm,right=2cm]{geometry}
\usepackage{tikz}
\usepackage[pdftex, pdftitle={Article}, pdfauthor={Author}]{hyperref}
\usepackage{lscape}

\begin{document}

 \vspace{1.5cm}
\begin{center}
\ \\
{\bf\Large Electron and monopole properties in a dark matter model}

\vspace{0.5cm}

Jacob Garc\'{\i}a-Moreno 

and

Vicente Vento \vskip 0.7cm

{\it  Departamento de F\'{\i}sica Te\'orica and Instituto de
F\'{\i}sica Corpuscular}

{\it Universidad de Valencia - Consejo Superior de Investigaciones
Cient\'{\i}ficas}

{\it 46100 Burjassot (Val\`encia), Spain, }

{\small Email: vicente.vento@uv.es}

\end{center}
\vskip 1cm \centerline{\bf Abstract}
Monopoles have been a subject of much theoretical and experimental research since they were proposed  to symmetrize Maxwell's equations.
No experimental signature of their existence has been detected. A model generalizing QED with dark massless photons and dark monopoles was proposed not long ago. In this model, by introducing a mixing interaction between the two sectors, leptons and monopoles are transformed  into dyons. This particular behavior of the electrons determines a new path to study the physics of monopoles. Here we generalize the Electroweak theory in the same way by incorporating a dark sector with massive photons and monopoles, and analyze experimental consequences for electrons in both models. The mass of the dark photon plays an important role regarding observability.

\vspace{1cm}

\noindent Pacs: 12.20-m,12.15.-y, 13.40.Hq,14.80.Hv

\noindent Keywords: monopoles, electroweak interactions, dark matter

\vspace{1cm}

\section{Introduction}

Dirac's beautiful interpretation of the quantum mechanical behavior of the magnetic monopole led to charge quantization\cite{Dirac:1931qs,Dirac:1948um}.  Later the development of 't Hooft and Polyakov \cite{tHooft:1974kcl,Polyakov:1974ek} proving that GUT theories contain monopoles gave a strong impulse for the search of monopoles \cite{Zeldovich:1978wj,Milton:2006cp,Acharya:2014nyr,Patrizii:2015uea,Mitsou:2026zcf}. However, those monopoles are extremely massive and mechanisms have been invented to describe lower mass monopoles \cite{Mavromatos:2020gwk}. Here, we consider a two photon theory for standard model particles and point-like monopoles, where the mass of the monopole is a parameter. The theory consists of a the conventional photon, which  interacts only with electric currents, and a dark photon interacting only with magnetic currents\cite{Brummer:2009oul,Terning:2018lsv}.  We next assume that the dark and ordinary sectors  interact through a portal in a manner that, though feeble, is experimentally accessible. In our portal the interaction takes place because of the kinetic mixing between the dark and the ordinary photons \cite{Holdom:1985ag,Fabbrichesi:2020wbt}.  Initially, we will generalize QED by adding  an additional $U(1)$ dark photon, and later on we will study a similar generalization for the Electroweak theory.  In the most general scenario of these models,  conventional particles and monopoles become dyons, acquiring electric and magnetic charges. In this study  our investigation deals with the determination of the mixing coupling constant. Once this coupling is determined  we  describe an experimental setup  where the peculiar behavior associated with the dark world arises. However, our study also produces a plethora of new observables, that will be discussed but their detailed analysis is left for future work.  

 \section{Dark QED}
 \label{QED}
Let us start with a two photon theory whose kinetic part is given by

\begin{equation}
\mathcal{L^\prime}_0 = - \frac{1}{4} \mathcal{F}_{\mu \nu} \mathcal{F}^{\mu \nu} - \frac{1}{4} \mathcal{G}_{\mu \nu} \mathcal{G}^{\mu \nu}\,,
\label{photons}
\end{equation}
where $ \mathcal{F}_{\mu \nu}  = \partial_\mu \mathcal{A}_\nu -\partial_{\nu} \mathcal{A}_\mu $ and  $\mathcal{G}_{\mu \nu}  = \partial_\mu \mathcal{C}_\nu -\partial_{\nu}\mathcal{C}_\mu $.
The new photon and  the monopoles form the  dark matter sector. Thus, $\mathcal{A}_\mu$ is the ordinary photon that couples only to particles with electric charge and  $\mathcal{C}_\mu$ the extraordinary photon that couples only to particles with magnetic charge. The interaction lagrangian with matter is 
 
 \begin{equation}
 \mathcal{L^\prime}_i = e J^e_\mu \mathcal{A}^\mu + g J^m_\mu \mathcal{C}^\mu\,.
 \label{matter}
 \end{equation}
 where $J^e_\mu $ is an electric current and $J^m_\mu $ a magnetic current. For simplicity we will consider ordinary matter to be made just of electrons and dark matter just of monopoles. We assume  both photons to be massless. In order to generate a portal we assume a kinetic mixing between the photons
 
 \begin{equation}
 \mathcal{L^\prime}_p= -\frac{\varepsilon}{2} \mathcal{F}_{\mu \nu} G^{\mu \nu}\,.
 \end{equation}
 This kinetic term can be diagonalized by transforming the photons \cite{Holdom:1985ag,Fabbrichesi:2020wbt}
 
 \begin{eqnarray}
{ \mathcal{C}^{\mu} \choose  \mathcal{A}^{\mu} } = \left( \begin{array}{cc}{\displaystyle \frac{1}{\sqrt{1-\varepsilon^2}}} &  0 \\
{\displaystyle -\frac{\varepsilon}{\sqrt{1-\varepsilon^2}}} & 1  \end{array} \right)  \left( \begin{array}{cc} \cos \theta & -\sin \theta \\
\sin \theta & \cos \theta  \end{array} \right) 
{ C^{ \mu} \choose  A^{\mu} }  \,,\label{rot}
\end{eqnarray}
 where now we identify $A^\mu$ with the {\it visible} photon and $C^\mu$ with the  {\it dark} photon.  We keep the nomenclature dark photon due to its origin despite the fact that it might couple weakly to ordinary particles as we explain next.
 
 After this transformation the photon lagrangian becomes diagonal
 
 \begin{equation}
{ \mathcal L}_0 = - \frac{1}{4} F_{\mu \nu} F^{\mu \nu} - \frac{1}{4} G_{\mu \nu} G^{\mu \nu}\,,
\end{equation}
where $ F_{\mu \nu}  = \partial_\mu A_\nu -\partial_{\nu} A_\mu$ and  $G_{\mu \nu}  = \partial_\mu C_\nu -\partial_{\nu} C_\mu.$ 
The interaction lagrangian Eq.(\ref{matter}) becomes \cite{Vento:2025tql}

\begin{eqnarray}
\mathcal{L}_i &=& \left( g J^m_\mu\frac{\cos \theta}{\sqrt{1-\varepsilon^2}} + eJ^e_\mu \left( \sin \theta - \frac{\varepsilon \cos \theta}{\sqrt{1-\varepsilon^2}}\right) \right) C^\mu \nonumber \\
& &  + \left( -g J^m_\mu\frac{\sin \theta}{\sqrt{1-\varepsilon^2}} + eJ^e_\mu \left( \cos \theta + \frac{\varepsilon \sin \theta}{\sqrt{1-\varepsilon^2}}\right) \right) A^\mu\,. 
\label{lagrangianQED}
\end{eqnarray}
This equation is very illuminating. It shows that in the transformed theory the electron becomes a dyon with electric and magnetic charge 

\begin{equation}
\left( e \left( \cos \theta + \frac{\varepsilon \sin \theta}{\sqrt{1-\varepsilon^2}}\right) , e \left( \sin \theta - \frac{\varepsilon \cos \theta}{\sqrt{1-\varepsilon^2}}\right)\right)\,,
\label{electron}
\end{equation}
and so does the monopole

\begin{equation}
\left(-g \frac{\sin \theta}{\sqrt{1-\varepsilon^2}},g \frac{\cos \theta}{\sqrt{1-\varepsilon^2}}\right)\,.
\label{monopole}
\end{equation}

Thus the dark photon gets a coupling to the electron and the ordinary photon to the monopole thanks to the mixing term. The new couplings depend not only on the parameter $\varepsilon$, a coupling constant, but also on the parameter $\theta$. For $\sin \theta = 0$ the ordinary photon must couple to the electron as in ordinary QED. This  fixes the normalization of the fields and therefore different values of $\theta$ lead to different charges, thus to different physical scenarios \cite{Holdom:1985ag,Fabbrichesi:2020wbt}.

 The Schwinger quantization condition for electron-monopole leads to \cite{Schwinger:1975ww}
 
 \begin{equation}
 \frac{e g}{\sqrt{1-\varepsilon^2}} =n ,\;\; n \in N\,.
 \label{schwinger}
 \end{equation}
 
 We are going to use next the softer scenario in which $\theta$ and $\varepsilon$ are small and therefore the charges for the electron and the monopole become

\begin{align}
&(e, -\varepsilon e) \;\; \text{for the electron,} \nonumber \\
&(0,\;\;\;\; g) \;\; \text{for the monopole.}
\label{charges}
\end{align}

\section{Determination of the kinetic mixing}

\begin{figure}[htb]
    \centering
    \includegraphics[scale=0.3]{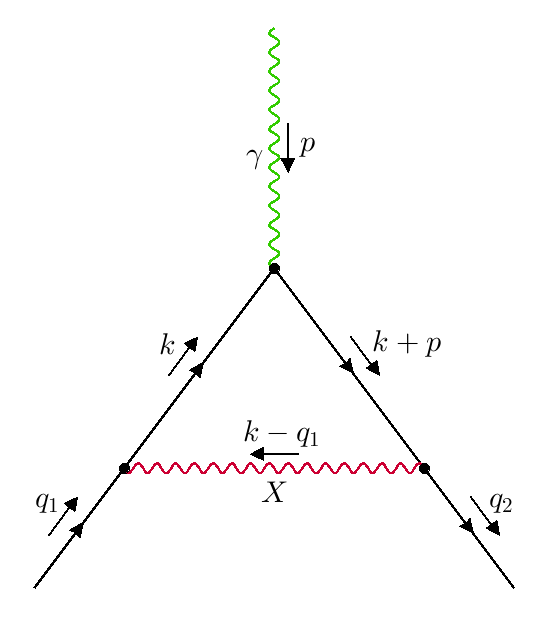}
    \caption{Feynman diagram associated with the lowest order contribution of the dark photon to the anomalous magnetic moment of the  electron.}
    \label{g-2}
\end{figure}

A way of detecting indirectly the existence of the dark photon and the monopoles is through the determination of the anomalous magnetic moment of the electron. We could easily generalize the model to all charged leptons, but given the fact that the electroweak and hadronic contributions to the electron are smaller than to the muon we will stick to the electron \cite{Raggi:2015yfk}. Given our lagrangian Eq.(\ref{lagrangianQED}) the interaction of the electrons with the dark photon assuming $\varepsilon$ small and $\theta \sim 0$  \cite{Vento:2025tql} becomes

\begin{equation}
-e\varepsilon\Bar{\psi} \gamma^\mu \psi X_\mu,
\end{equation}
where $\psi$ is the electron spinor. Thus, the contribution of the dark photon to $a_e = (g_e-2)/2 $ at lowest order is described by the Feynman diagram shown in Fig. \ref{g-2}.

In the present case where the dark photon has no mass the calculation is similar to that of conventional QED leading to

\begin{equation}
a_e^{dp} = \frac{\alpha}{2\pi}\varepsilon^2.
\end{equation}
where $\alpha$ is the fine structure constant.

The difference between the theoretical value calculated  with QED \cite{Aoyama:2012wj}

\begin{equation}
a_e^{th} = 1.15965218161(23),
\end{equation}
and the measured experimental value \cite{VanDyck:1987ay,Hanneke:2008tm}

\begin{equation}
a_e^{exp} =1.15965218159(13),
\end{equation}
is

\begin{equation}
    \Delta a_e = a_e^{exp} - a_e^{th} = -2 (26)\cdot 10^{-12}.
    \label{delta}
\end{equation}

If we assume the maximum difference between the experimental value and the theoretical value as the contribution from the dark photon  we are able to estimate the maximum posible value for the kinetic coupling $\varepsilon$ by using a constant value for the fine structure constant obtained from ref.  \cite{Parker:2018vye}.

\begin{equation}
  a_e^{dp} = \frac{\alpha}{2\pi}\varepsilon^2  = a_e^{QED} \varepsilon^2 , 
\end{equation}

which leads to

\begin{equation}
\varepsilon = \sqrt{\frac{2\pi \Delta a_e}{\alpha}} < 14 \cdot 10^{-5}.
\end{equation}

Despite our previous comments and for consistency lets us check the result for the anomalous magnetic moment of the muon. In this case  the difference between the theoretical value \cite{Aoyama:2012wk,Aoyama:2020ynm,Aliberti:2025beg,Boccaletti:2024guq}  and the experimental value \cite{Muong-2:2006rrc,Muong-2:2025xyk}  is

\begin{equation}
    \Delta a_\mu = a_\mu^{exp} - a_\mu^{th} = 134.2(16.6)\cdot 10^{-11}.
    \label{eq:dif_momento}
\end{equation}
which leads to  $\varepsilon < 113 \cdot 10^{-5}$, a larger value which is a consequence of the larger difference between the experimental and the theoretical values. However, both experimental analysis lead to a maximum value for the coupling constant of about $10^{-4}$.

\section{Biot-Savart experiment revisited}

\begin{figure}[htb]
\begin{center}
\includegraphics[scale= 0.3, angle=270]{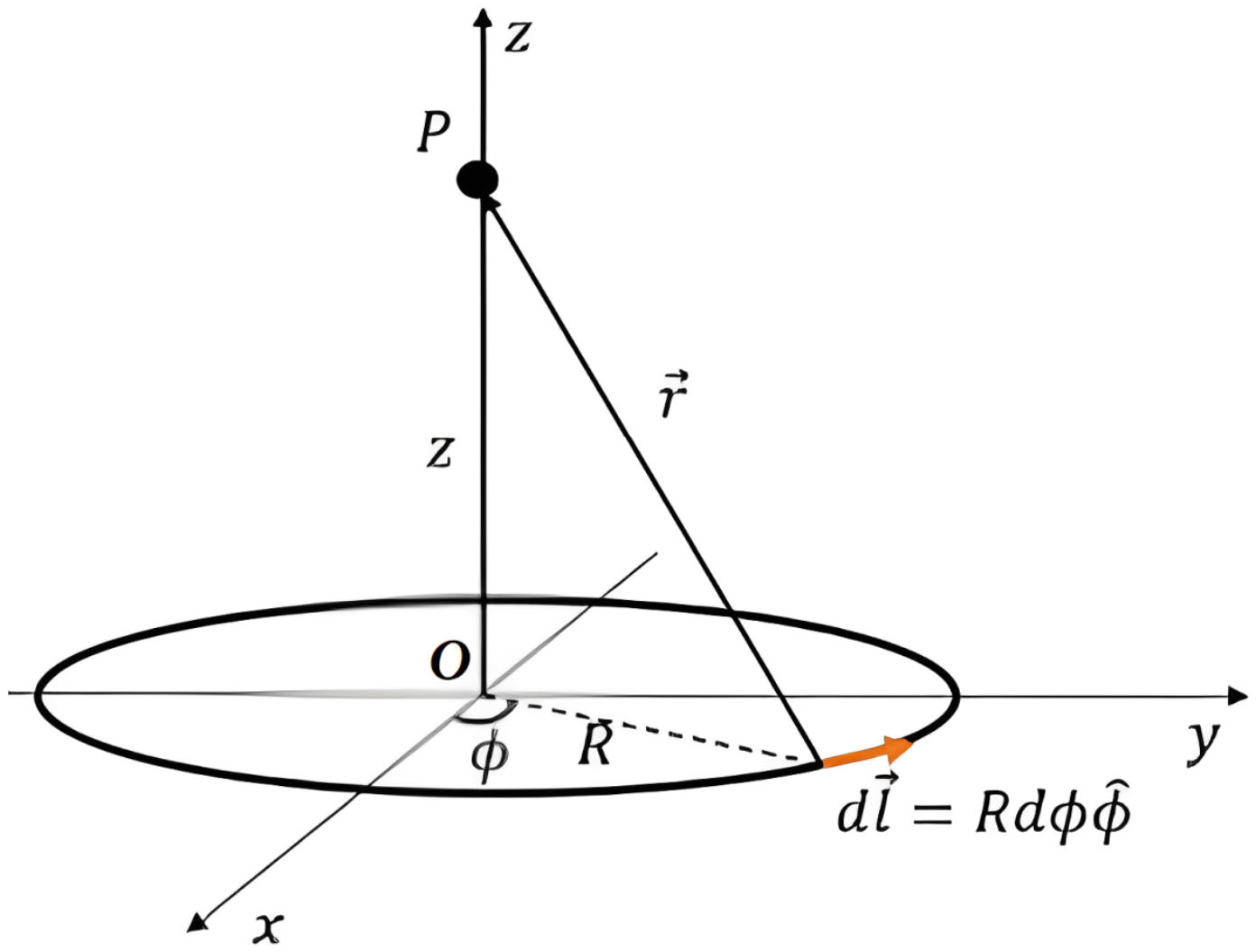} 
\caption{Circular coil of the the Biot-Savart experiment.}
\label{coil}
\end{center}
\end{figure}

Having calculated the coupling associated with the kinetic mixing,  we know, Eq.(\ref{charges}), not only the electric but also the magnetic charge of the electron. Let us now study some peculiar consequence of the magnetic charge of the electron. A classic experiment that will show the effect of the magnetic charge is the Biot-Savart experiment \cite{Jackson:1998nia}.  In Fig. \ref{coil} we show the experimental setup with notation.

\begin{figure}[htb]
    \centering
    \includegraphics[scale=0.55]{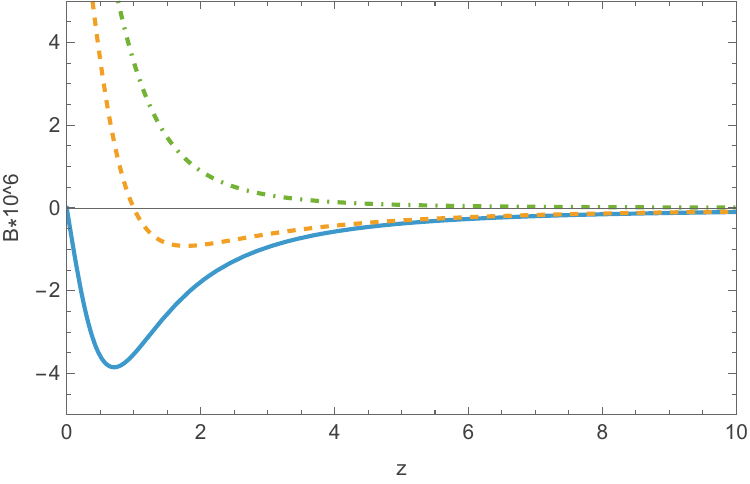} \hspace{0.5cm}  \includegraphics[scale=0.55]{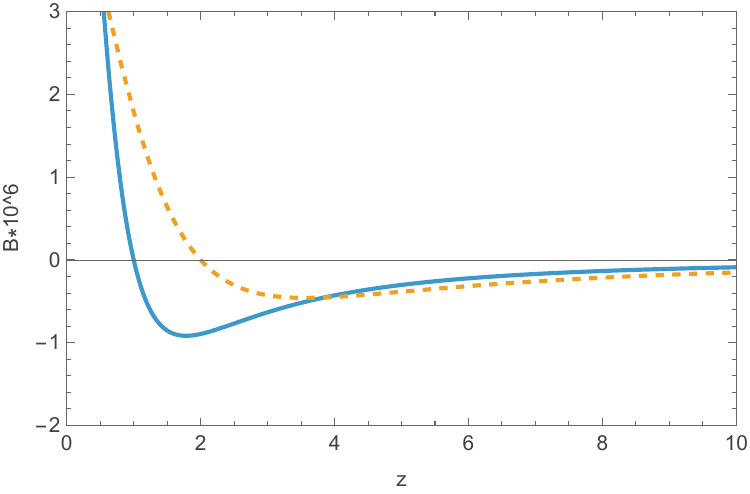} \\
     \includegraphics[scale=0.55]{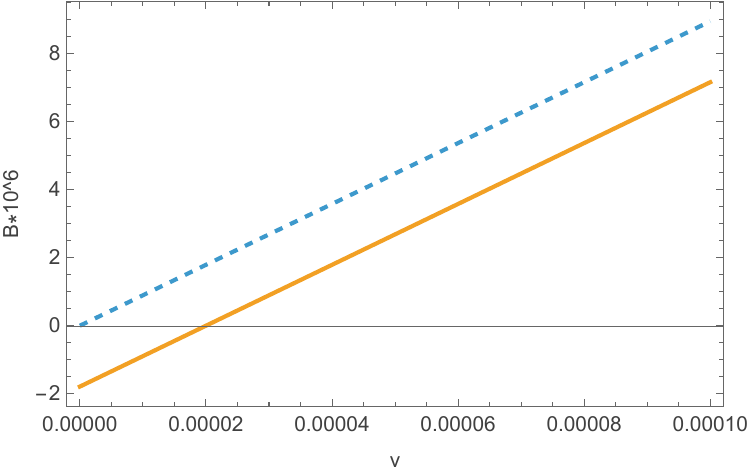} 
 \caption{We plot the magnetic field B in units of $2\pi e n A/R$ : i)  left figure : as a function of $z=Z/R$  for $\varepsilon=10^{-5}$ and $v=0$ (solid); for   $\varepsilon=10^{-5}$ and $v=10^{-5}$ (dashed); for $\varepsilon=0$, and $v=10^{-5}$ (dotdashed); ii) right figure: as a function of $z$ for $\varepsilon=10^{-5}$, $v=10^{-5}$ and a  fixed $R$ (solid); and $\varepsilon=10^{-5}$, $v=10^{-5}$ and double the radius (dashed); iii)  figure below: as a function of $v$ for $\varepsilon=0$, and $z=2$ close to the maximum of the right figure (dashed) and $\varepsilon=10^{-5}$ and $z=2$ (solid).}
    \label{BS}
\end{figure}

Recalculating the classic Biot-Savart experiment in the presence of a rotating electron current with not only electric but also a magnetic charges, the result for small velocities becomes \cite{Vento:2025tql}. 

\begin{equation}
B  = \left(1+ \frac{Z^2}{R^2}\right)^{-\sfrac{3}{2}} \left( \frac{v}{c} - \varepsilon \frac{Z}{R}\right) \frac{2\pi e n A}{R}.
\label{BR}
\end{equation}
In Fig. \ref{BS} we plot the magnetic field for these electrons as the outcome of such an  experiment, analyzing the sensibility to the mixing parameter $\varepsilon$ in the range just obtained, and to the electron velocity $v$  and to the vertical distance  $Z$ and radius of the coil $R$ of the experiment. In the figure we use $z=Z/R$ and $v= v/c$. 

The figure shows clearly the characteristics of the experiment. The region between $z=1$ and $z=3$ is the most important. There, as can be seen in the left figure, in the appropriate conditions, the magnetic field will vanish and change sign., i.e. turn counterclockwise if without magnetic charge it should turn clockwise. This change of direction implies the existence of a magnetic charge. In the right figure we study the variation of the effect with the size of the circle: the bigger the radius, the smaller the effect. Finally in the figure below we chose one value of $z$ close to the maximum and we study the effect of the velocity. We see that the smaller the velocity the bigger the effect, as appears natural from Eq.(\ref{BR}).  Evidently, we can also see that the larger $\varepsilon$ the bigger the effect. Finally in order to asses the measurability of the effect one has to look at the units $2\pi e n A/R$ it is clear that the effect will become bigger for higher densities $n$, thicker cables $A$ and smaller radii $R$.

\section{Dark Electroweak Theory}
Let us extend the study presented in section \ref{QED} to the Electroweak Theory.   In our case we add to the conventional  $SU_L(2)\otimes U_Y(1)$ theory a dark photon $X$ via an additional abelian group  $U_X(1)$ \cite{Sun:2023kfu,Jgm:2026tfm}. Other extensions have been discussed in the literature \cite{Baek:2013dwa}. The  kinetic term of this $SU_L(2)\otimes U_Y(1) \otimes U_X(1)$ theory is

\begin{equation}
    \mathcal{L}_0 = -\frac{1}{4}\Tilde{X}'_{\mu\nu}\Tilde{X}'^{\mu\nu} - \frac{\varepsilon}{2c_W}\Tilde{X}'_{\mu\nu}\Tilde{Y}'^{\mu\nu} - \frac{1}{4}\Tilde{Y}'_{\mu\nu}\Tilde{Y}'^{\mu\nu} -\frac{1}{4}\Tilde{W}'^a_{\mu\nu}\Tilde{W}'^{a,\mu\nu},
 \label{kinetic1}
\end{equation}
where the index $a$ runs over $SU_L(2)$, and $Y$ represents the hypercharge.  We introduce a kinetic coupling between the dark photon and in this case the  $Y$ boson, and for convenience the coupling has been defined in terms of the Weinberg angle. We use the conventional abbreviated notation for the transcendental functions ($sin \rightarrow s, cos \rightarrow c, tan \rightarrow t$).  

The interaction term for electrons, neutrinos and monopoles is 

\begin{align}
    \mathcal{L}_i =& -\frac{1}{2}\Bar{e}_L\gamma^\mu e_L (g'\Tilde{Y}'_\mu + g \Tilde{W}'^3_\mu) - \Bar{e}_R\gamma^\mu e_Rg'\Tilde{Y}'_\mu + \frac{g}{\sqrt{2}}\Bar{\nu}_L\gamma^\mu e_L \Tilde{W'}_\mu^+ +  \frac{g}{\sqrt{2}}\Bar{e}_L\gamma^\mu \nu_L \Tilde{W'}_\mu^- \nonumber \\
    &+ \frac{1}{2}\Bar{\nu}_L \gamma^\mu \nu_L(g\Tilde{W'}_\mu^3 - g'\Tilde{Y}'_\mu) + g_XJ_X^\mu \Tilde{X}'_\mu,
     \label{interaction1}
\end{align}
where $g_X$ is the coupling for the dark photon and $J_X$ represents the monopole current.

 One proceeds by diagonalizing the fields to one obtain a theory where a massive dark photon is coupled not only to the monopole but also to conventional particles \cite{Sun:2023kfu,Jgm:2026tfm}. To this theory we incorporate Higgs fields  to endow masses to the heavy bosons and the dark photon. In the approximation  that the coupling between the dark world and the ordinary world  is weak the angle $\delta$ 
 \begin{equation}
    \begin{pmatrix}
        \Tilde{Z}_\mu \\
        \Tilde{X}_\mu
    \end{pmatrix} = \begin{pmatrix}
        c_\delta & -s_\delta \\
        s_\delta & c_\delta
    \end{pmatrix} \begin{pmatrix}
        Z_\mu \\
        X_\mu
    \end{pmatrix},
    \label{masa}
\end{equation}
 describing the coupling of the $Z$ and $X$ bosons  is small, i.e,  $s_\delta \sim \varepsilon $ and $c_\delta \sim 1$, and therefore the corrections to couplings of the $Z$ boson to the conventional particles  are negligible $\sim \varepsilon^2$, and therefore they remain the same as that of the standard model. Thus in  this model the masses of the weak bosons are the same as in the standard model, and the dark photon acquires a mass $m_X$. 
 
 Summarizing, the theory has besides a conventional kinetic terms, a kinetic term for a massive dark photon,  and the following interaction lagrangian for one lepton family \cite{Jgm:2026tfm} 

\begin{align}
    \mathcal{L}_i &= \frac{g}{\sqrt{2}}\Bar{\nu}_L\gamma^\mu e_L W_\mu^+ + \frac{g}{\sqrt{2}}\Bar{e}_L\gamma^\mu \nu_L W_\mu^- -eJ_e^\mu A_\mu+e\varepsilon J_e^\mu X_\mu + g_X J_X^\mu\left[\delta Z_\mu +X_\mu\right]  \nonumber \\
    &\frac{g}{2c_W}J_\nu^\mu Z_\mu+\frac{gt_W\varepsilon}{2c_W}\left(\frac{\omega^2+\sigma}{1+\omega^2}\right)J_\nu^\mu X_\mu + J_e^\mu Z_\mu (g_V - g_A\gamma^5) +\varepsilon t_W  \left(\frac{\omega^2+\sigma}{1+\omega^2}\right)J_e^\mu X_\mu(g_V -g_A\gamma^5).
 \label{lagrangianEW} 
\end{align}
This interaction lagrangian modifies  Eq.(\ref{lagrangianQED}) to incorporate the weak interaction. Here $\delta = \frac{\varepsilon t_W(1-\sigma)}{1+\omega^2}$ , $\omega^2= \frac{m_X^2}{m_Z^2}$, $\sigma= 2 \frac{g_Xc^2_W}{e \varepsilon} \frac{v_2^2}{v^2}$ , where $v$ corresponds to the vacuum expectation value $v=246$  GeV, and $v_2$ is an additional parameter of the theory associated with the Higgs mechanisms\cite{Jgm:2026tfm}. Here $J_e^\mu$ represents the electron current, $J_\nu^\mu$ the neutrino current and $J_X^\mu$ the monopole current. Note that we have kept a term $g_X \delta \sim g_X \varepsilon$, which we consider not negligible due to the large value of the monopole coupling constant. The Higgs sector is of no interest in our model.

In this regime of weak coupling and small $\delta$ we have the following interesting new interactions.  The electron acquires a small magnetic charge $e\varepsilon$ as before. New interactions between electrons and neutrinos with the dark photon appear with the shape of neutral currents due to the coupling with the $Z$. A new interaction for the monopole mediated by $Z$ that is controlled by the mass coupling of the $Z$ and the dark photon coupling $\sigma$.  The new interactions depend strongly on the mixing coupling constant $\varepsilon$. We saw that it was very small in pure QED, and if the same happens here the effect of these new interactions will be small. Therefore is most important to see if the addition of mass to the dark photon changes our expectations. 

\section{Determination of the kinetic mixing}

The most important contribution to the  calculation of the electron magnetic moment  in the model is that shown in Fig. \ref{g-2} where now the dark photon has mass.  The result is

\begin{equation}
a_e^{dp} = \frac{\alpha}{\pi} \varepsilon^2 \int^1_0 \frac{z(1-z)^2}{(1-z)^2+z \omega_m},
\end{equation}
where $\omega_m=m_X^2/m^2$ is a mass ratio of the photon to the electron masses. The exact result is non analytic but it can be shown that for $m_X<<m$ we get the results of the massless model and for $m_X>>m$ we obtain $a_e^{dp} = \frac{2}{3} \frac{m^2}{m_X^2} \varepsilon^2  a_e^{QED}$. Note that for large masses the $a_e^{dp}$ is directly proportional to $\varepsilon^2$ and inversely proportional to $m_X^2$. This dependence is crucial to understand the behavior of the limiting coupling $\varepsilon$.

\begin{figure}[htb]
    \centering
    \includegraphics[scale=0.55]{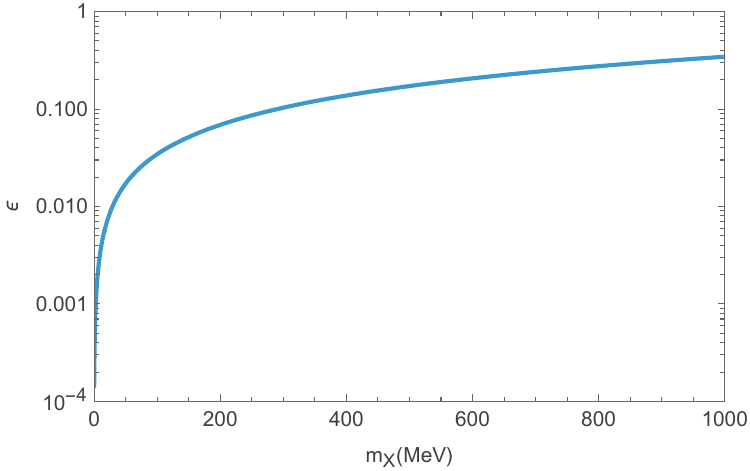}  \hspace{0.5cm}  \includegraphics[scale=0.55]{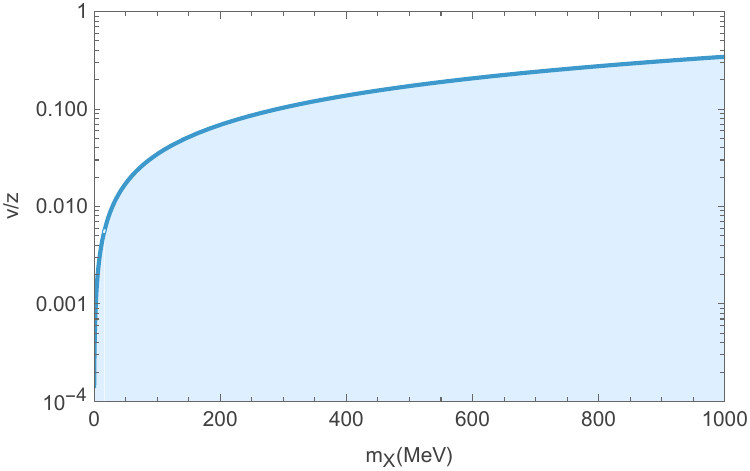} \
    \caption{Figure left: We plot the the value of the dark coupling constant as a function of the the dark photon mass. Figure right: for fixed $z$ the shaded region represents the velocities for which the magnetic field $B$ changes is rotation sense due to the magnetic charge of the electron. }
    \label{coupling}
\end{figure}

We proceed by making again the assumption that the precision of the measurement of $g-2$ determines the contribution of the dark photon to the magnetic moment. This assumption allows us to establish a relation between the dark coupling parameter $\varepsilon$ and the dark photon mass, which we show in Fig. \ref{coupling}. By looking at the figure one concludes  that the maximum possible value for the mixing coupling constant $\varepsilon$  increases with the mass of the dark photon, leading thus to more observable effects.

\section{Biot-Savart  experiment for massive dark photons}

Let us repeat the Biot-Savart experiment with the functional form of the mixing coupling $\varepsilon$. As we saw in Fig. \ref{BS} a clear experimental signature of our dark theories in the case of the electron is the fact that the magnetic field reverses its rotation in the Biot-Savart experiment. From Eq.(\ref{BR}) it is clear that the sign changes  for  $v/c < \varepsilon Z/R$. The main effect on the electron of having  provided the dark photon with a mass is the strong variation shown in the allowed coupling constant. In the right side of Fig. \ref{coupling}  we show how this functional form affects the interesting electron velocities which allow in the Biot-Savart experiment the reversing of the magnetic field to take place. The possible velocities become larger for massive dark photons making the experiment easier to observe.

\begin{figure}[htb]
\centering
  \includegraphics[scale=0.6]{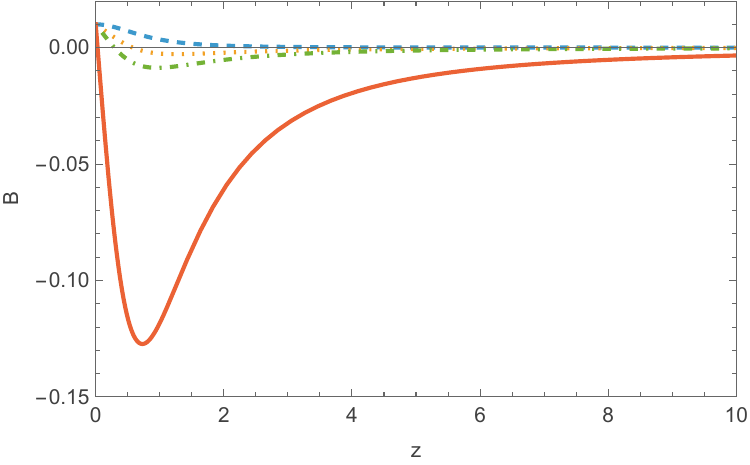} 
  \vskip - 3.7cm 
  \hskip 3.2cm 
   \includegraphics[scale=0.3]{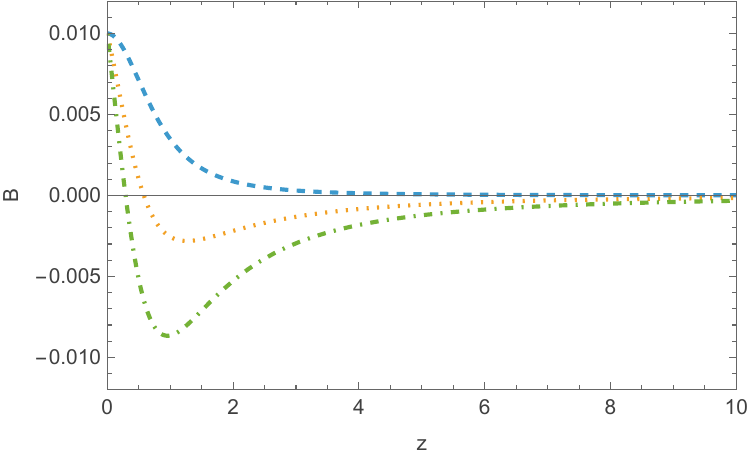} 
     \vskip 1.2cm
    \caption{We plot the magnetic field $B$ of a Biot-Savart experiment for electrons with magnetic charge  in units of $2\pi e n A/R$ as a function of $z$ for different dark photon masses: $m_X=0 \; \text{(dashed)}$ ; $50  \text{ MeV (dotted)}$ ; $100 \text{ MeV (dotdashed)}$ ; $1000 \text{ MeV (solid)}$. The inset enlarges the plot for the lower masses.}
    \label{BRmX}
\end{figure}

In order to see the interesting outcome of the experiment we show in Fig. \ref{BRmX}   the magnetic field $B$ for $v=0.01, z=2,$ and $m_X= 0,50,100,1000$ MeV. In this way it is apparent  the effect of increasing the coupling. We notice that the effect in the Biot-Savart experiment occurs for much bigger velocities as compared to the massless case and produces much bigger, and therefore easier to observe, inverse magnetic fields. 

The impact of the dark photon mass on the observables is significant. As the mass increases, the achievable coupling constant also grows, and the corresponding velocities and magnetic fields at which the Biot-Savart effect manifests itself become higher. Furthermore, the possibility of a larger mixing coupling results in stronger couplings in the Electroweak interaction Lagrangian of Eq.(\ref{lagrangianEW}), thereby enabling a range of interesting phenomena: monopole production from dark photons or $Z$, dark photon contributions to the $Z$ width, dark photon contributions to $e^+e^-\rightarrow \mu^+ \mu^-$, and so on, phenomena that could potentially become observable.

\section{Monopole production}

\begin{figure}[htb]
    \centering
   \includegraphics[scale=0.35]{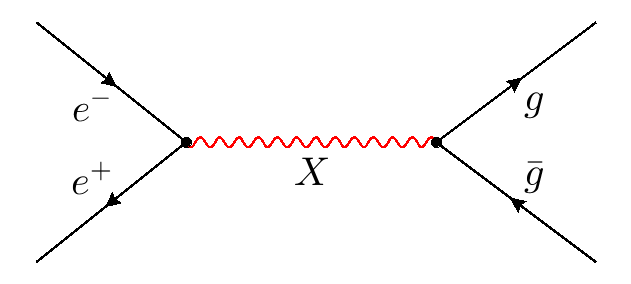}       
   \caption{Biggest lowest order contributions to the electron-positron production of monopoles}
    \label{eeXgg}
\end{figure}

As an example of other observations directly related to the model let us describe the electron-positron production of monopole-anti-monopole. In this calculation we are going to consider the dark photon as an elementary particle. The main contribution to this process according to our interaction lagrangian Eq.(\ref{lagrangianEW}) comes from the dark photon exchange, see Fig. \ref{eeXgg}, whose amplitude is proportional to $\varepsilon e g $, since in our approximation $g_X=g$ as shown in Eq.(\ref{charges}). Moreover, due to the quantization condition Eq.(\ref{schwinger}), $\varepsilon e g \sim \varepsilon$. A more complete calculation including the contribution of the smaller $Z$ exchange diagram will be published in the future including also the width of the dark photon as determined from the interaction lagrangian \cite{Jgm:2026tfm}.

Let us consider initially that the monopoles are spin 0 bosons and recall that the dark photon is a spin one massive particle, thus  a Proca field. The calculation of the differential cross section in the center-of-mass (CM) frame is straightforward and leads to

\begin{equation}
\frac{d\sigma}{d\Omega} = \frac{\varepsilon^2} {128 \pi^2 s}\frac{\beta_g^3}{\beta_e}\frac{\sin^2 \Theta}{(1-\frac{m_X^2}{s})^2}.
\end{equation}
Here $\Omega$ is the solid angle, $s$ the center of mass energy squared, $\beta_g = \sqrt{1- \frac{m_g^2}{4 s}}$, $\beta_e = \sqrt{1- \frac{m_e^2}{4 s}}$, where $m_g$, $m_X$ and $m_e$ are the monopole, dark photon  and electron mass respectively and  $\Theta$ the CM scattering angle. A peculiarity of the production of bosons, due to $\sin^2 \Theta$ in the differential cross section, is that the scattering is suppressed forward and backward, and therefore it peaks at $\Theta = \frac{\pi}{2}$, i.e. the scattering is p-wave.

The corresponding total CM cross section becomes

\begin{equation}
\sigma= \frac{\varepsilon^2}{48 \pi^2} \frac{\beta_g^3}{\beta_e} \frac{1}{(1-\frac{m_X^2}{s})^2}.
\end{equation}

\begin{figure}[htb]
    \centering
    \includegraphics[scale=0.55]{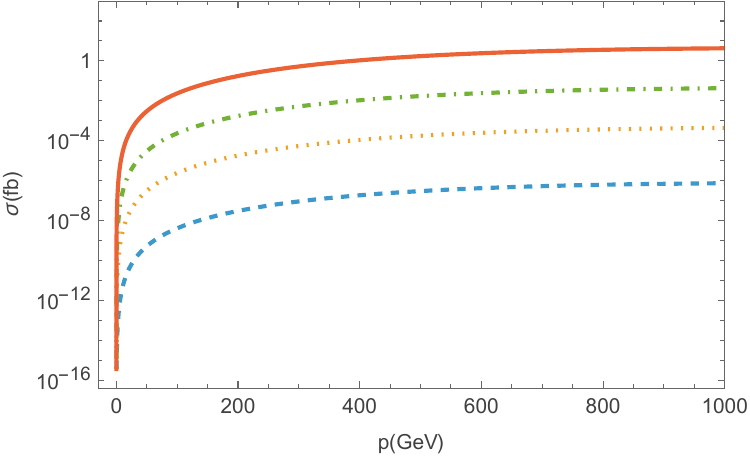} \hspace{0.5cm}  \includegraphics[scale=0.55]{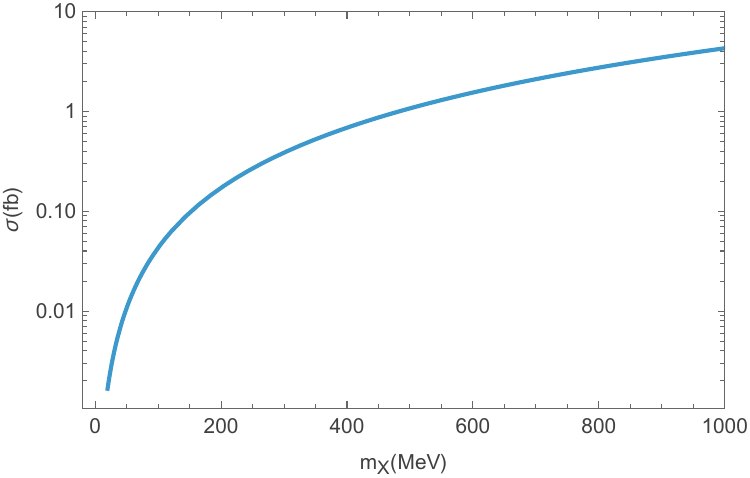} \\
     \includegraphics[scale=0.55]{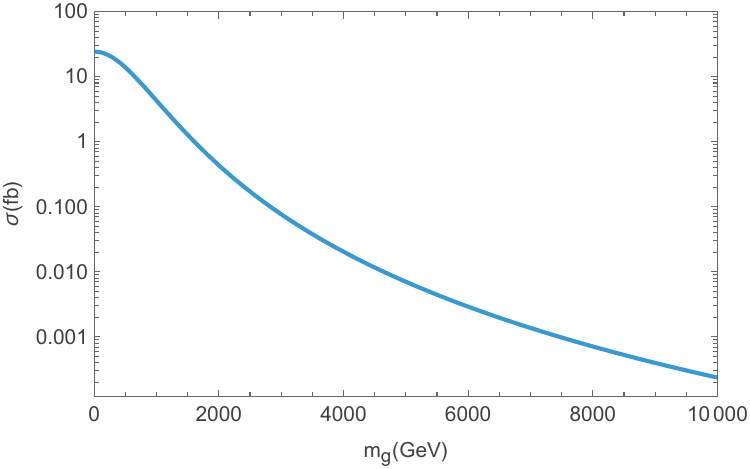} 
 \caption{Figure left: we plot the total CM cross section as a function of monopole momentum for fixed monopole mass $m_g=1$ TeV, and for different values of the dark photon mass $m_x= 0 \;\text{(dashed)}; 10  \text{MeV \; (dotted)}; 100  \text{MeV \; (dotdashed)}; 1000  \text{MeV \; (solid)}$. Figure right: we plot the total CM cross section as a function of dark photon mass $m_X$ for fixed monopole mass and momentum $m_g=p= 1 $TeV. Figure below: we plot the total center of mass cross section as a function of monopole mass $m_g$ for fixed dark photon mass $m_X=1$ GeV and fixed momentum $p=1$TeV.}
    \label{sigmab}
\end{figure}

Let us analyze the total cross section for different experimental scenarios. In Fig. \ref{sigmab} left we show the cross section in the CM as a function of momentum.  The cross section is very small for small momenta and tends to stabilize for momenta of the order of the monopole mass. We have taken $m_g = 1$ TeV for this analysis. We show in the graphic the cross section for dark photon masses up to $1$ GeV. Since $\varepsilon$ grows with the dark photon mass,  the cross section does too.  The cross sections obtained are compatible with present experimental limits \cite{Milton:2006cp,Patrizii:2015uea,Mitsou:2026zcf} . We plot in the right side of the figure the dependence on  $m_X$ for fixed $m_g=p=1$ TeV,  the maximum values within the flat region of the previous graphics. Here it is apparent the growth with $m_X$ as a consequence of the growth of $\varepsilon$. Finally below, we plot the dependence on monopole mass $m_g$ for $m_X= 1$ GeV and   $p= 1$ TeV. It becomes apparent that the cross section decreases as the mass of the monopole is increased.

\begin{figure}[htb]
    \centering
    \includegraphics[scale=0.55]{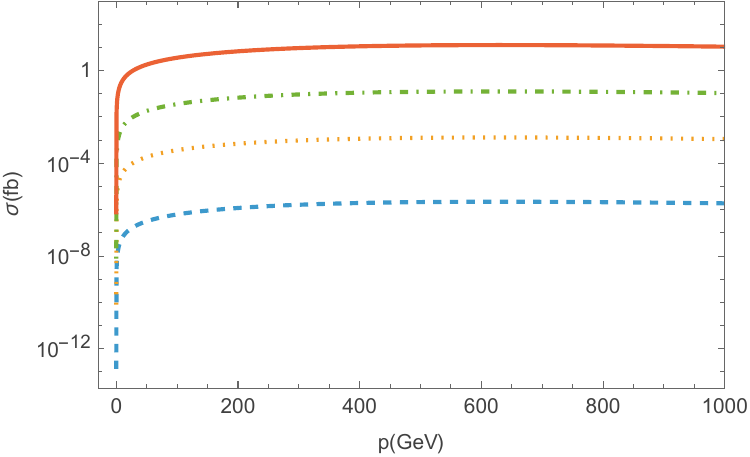} \hspace{0.5cm}  \includegraphics[scale=0.55]{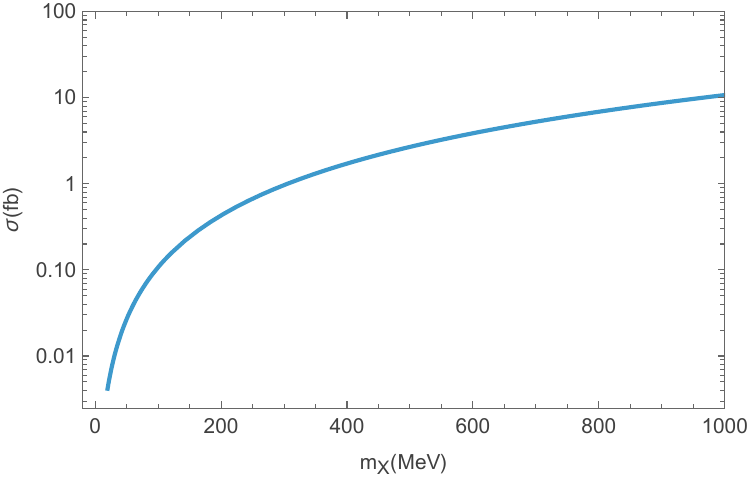} \\
     \includegraphics[scale=0.55]{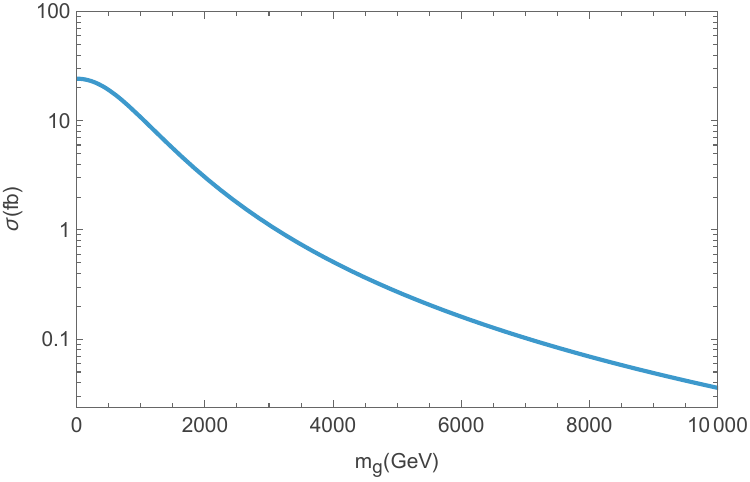} 
 \caption{Figure left: we plot the total CM cross section as a function of monopole momentum for fixed monopole mass $m_g=1$ TeV and for different values of the dark photon mass $m_x= 0 \;\text{(dashed)}; 10  \text{MeV \; (dotted)}; 100  \text{MeV \; (dotdashed)}; 1000  \text{MeV \; (solid)}$. Figure right: we plot the total CM cross section as a function of dark photon mass $m_X$ for fixed monopole mass and momentum $m_g=p= 1 $TeV. Figure below: we plot the total center of mass cross section as a function of monopole mass $m_g$ for fixed dark photon mass $m_X=1$ GeV and fixed momentum $p=1$TeV.}
    \label{sigmas}
\end{figure}

For spin 1/2 monopoles the main change in the differential cross section is the angular dependence which now becomes fundamentally proportional to $1+\cos^2\theta$, thus  maximizing forward scattering. This allows spinors  to have an  s-wave scattering while in the case of bosons it was in p-wave. Therefore, the suppression we noted at threshold due to the $\beta_g^3$ factor, in the cross section, now disappears because now the dependence goes like $\beta_g$ , 

\begin{equation}
\sigma= \frac{\varepsilon^2}{48 \pi^2} \frac{\beta_g^3}{\beta_e} \frac{1}{(1-\frac{m_X^2}{s})^2}\left( 1+ \frac{2m_e^2}{s} + \frac{2 m_g^2}{s}\right).
\end{equation}
Similar experimental scenarios as before are shown  in Fig. \ref{sigmas}.

The cross section is very small for small momenta but rises fast, due to the mentioned threshold behavior, to the stability plateau  Moreover, it tends to be larger for higher momenta, than for the boson case, due to the extra mass factors. However, qualitatively it is very similar and in both cases observability grows with the dark photon mass to observable scenarios. The behavior with $m_X$ although steeper is also qualitatively similar, as is the dependence with $m_g$. 

This calculation of monopole production, which has been performed completely within the dark sector, shows again that for large dark photon masses,  electroweak phenomena , including monopole physics, might be observable.

\section{Conclusions}
We have investigated a dark matter extension of QED featuring two massless photons, in which leptons acquire a magnetic charge, and examined the contribution of the dark photon to the anomalous magnetic moment ($g-2$) of leptons. This calculation provided a bound on the maximum mixing coupling constant. With this constraint in hand, we revisited the previously proposed Biot-Savart experiment to assess its observational feasibility.

We then extended the model to the electroweak theory and incorporated massive dark photons. The aim was to enlarge the Standard Model into the dark sector, consisting of a photon and monopoles. We introduced thereafter the appropriate Higgs mechanisms to generate masses for the heavy mesons and the dark photon. The resulting model exhibits a rich phenomenology. Within this framework, we computed the dark photon contribution to the electron $g-2$ and found that the mixing coupling, whose maximum value was extremely small in the massless photon case, grows substantially with the dark photon mass. As shown in Fig. \ref{coupling}, assuming that the coupling constant accounts for the unexplained portion of the $g-2$, a quadratic dependence on the photon mass emerges. We applied this new result to the Biot-Savart experiment, which becomes experimentally more feasible for reasonable dark photon masses. Clearly, a positive signal in that experiment, as discussed in detail in the massless case, would imply not only the existence of dark photons but also of monopoles.

To end it must be stressed that the Lagrangian of Eq.(\ref{lagrangianEW}) reveals new interactions between electrons and neutrinos with the dark photon, taking the form of neutral currents arising from the coupling to the $Z$. In addition, a new monopole interaction mediated by the $Z$ appears, controlled by the mass coupling $\sigma$ between the $Z$ and the dark photon. For large dark photon masses, these interactions will give rise to a rich phenomenology to be investigated in future work. In other to confirm that statement we have calculated the production of monopoles by electron-positron annihilation to lowest order both for spin 0 monopoles and spin 1/2 monopoles. The main result for this investigation is again the important effect of the dark photon mass, which leads to cross sections which are compatible with experimental non observation results, but close to the observation limit.

\section*{Acknowledgments}
VV acknowledges illuminating email exchanges  with D. Singleton regarding the two photon model, Marco Fabbrichesi regarding some aspects of ref. \cite{Fabbrichesi:2020wbt}, and Arcadi Santamaria regarding field redefinitions. Support by Grant CNS2022-135688 Consolidaci\'on Investigadora and SO project CEX2023-001292-S funded by MICIU/AEI are acknowledged.

\section*{Funding} Open Access funding provided thanks to the CRUE-CSIC agreement with Springer Nature.

\section*{Data Availability Statement} No Data associated in the manuscript.

\end{document}